\documentclass[aps,prl,twocolumn,preprintnumbers,superscriptaddress,amsmath]{revtex4}

\usepackage{amssymb}
\usepackage{txfonts}
\usepackage{amssymb}
\usepackage{graphicx}
\usepackage{subfigure}
\usepackage{dcolumn}
\usepackage{bm}
\usepackage{color}
\usepackage{upgreek}

\usepackage{tabularx}
\usepackage{makecell}

\begin{document}

\title{Silicon photonic devices for scalable quantum information applications}

\author{Lantian Feng}
\affiliation
{CAS Key Laboratory of Quantum Information, University of Science and Technology of China, Hefei 230026, China}
\affiliation{CAS Center for Excellence in Quantum Information and Quantum Physics, University of Science and Technology of China, Hefei 230026, China}
\author{Ming Zhang}
\affiliation{State Key Laboratory for Modern Optical Instrumentation, Centre for Optical and Electromagnetic Research, Zhejiang Provincial Key Laboratory for Sensing Technologies, Zhejiang University, Zijingang Campus, Hangzhou, 310058, China}
\author{Jianwei Wang}
\affiliation{State Key Laboratory for Mesoscopic Physics, School of Physics, Peking University, Beijing, China}
\affiliation{Frontiers Science Center for Nano-optoelectronics, Collaborative Innovation Center of Quantum Matter, Peking University, Bejing, China}
\author{Xiaoqi Zhou}
\affiliation{School of Physics and State Key Laboratory of Optoelectronic Materials and Technologies, Sun Yat-sen University, Guangzhou, 510000, China}
\author{Xiaogang Qiang}
\affiliation{National Innovation Institute of Defense Technology, AMS, 100071 Beijing, China}
\author{Guangcan Guo}
\affiliation
{CAS Key Laboratory of Quantum Information, University of Science and Technology of China, Hefei 230026, China}
\affiliation{CAS Center for Excellence in Quantum Information and Quantum Physics, University of Science and Technology of China, Hefei 230026, China}
\affiliation{Hefei National Laboratory, University of Science and Technology of China, Hefei 230088, China}
\author{Xifeng Ren}
\affiliation
{CAS Key Laboratory of Quantum Information, University of Science and Technology of China, Hefei 230026, China}
\affiliation{CAS Center for Excellence in Quantum Information and Quantum Physics, University of Science and Technology of China, Hefei 230026, China}
\affiliation{Hefei National Laboratory, University of Science and Technology of China, Hefei 230088, China}
\affiliation{e-mail: renxf@ustc.edu.cn}

\begin{abstract}
With high integration density and excellent optical properties, silicon photonics is becoming a promising platform for complete integration and large-scale optical quantum information processing. Scalable quantum information applications need photon generation and detection to be integrated on the same chip, and we have seen that various devices on the silicon photonic chip have been developed for this goal. This paper reviews the relevant research results and state-of-the-art technologies on the silicon photonic chip for scalable quantum applications. Despite the shortcomings, properties of some components have already met the requirements for further expansion. Furthermore, we point out the challenges ahead and further research directions for on-chip scalable quantum information applications.
\end{abstract}
\maketitle

\section*{I. INTRODUCTION}
Quantum information science is a new frontier subject combining quantum mechanics and information science. The quantum nature of particle superposition, entanglement, and measurement are applicable for more efficient information processing, computation, transmission and storage. The development of quantum information industry promises more powerful computing power, high security information communication and a deeper understanding of nature. It is expected to resolve numerous scientific problems that are difficult to be effectively solved by existing classical techniques. For example, absolutely secure information transmission \cite{Bennett1984,Gisin2002}, exponentially or polynomial accelerating the resolving of hard scientific problems \cite{Shor1994,Grover1997}, efficient simulation of molecular structures \cite{Feynman1982} and precision measurement beyond the standard quantum limit \cite{Giovannetti2011}.

In the last decades, quantum fundamental science is rapidly transformed into quantum technologies with huge resources invested by global academia, research centers and industry. Quantum scientific research is moving from the stage of principle verification of quantum rules to the stage of practical device research and development governed by these rules. We have seen that quantum computational advantage realized in different systems \cite{Arute2019,Zhong2020,Wu2021}, quantum key distribution via thousands of kilometer level satellite-to-ground quantum network \cite{Liao2017} and quantum-enhanced sensing for gravitational waves \cite{Abbott2016}, magnetic field \cite{Boss2017} and mass of nanoparticle \cite{Zheng2020}.

Photon, as one ideal information carrier, has been widely used in quantum information processing and shows several unique advantages such as fast transmission speed, low noise, multiple degrees of freedom for information encoding and high capacity. Besides, light has a wide range of applications in energy, communication, computation, medical care etc. These increasingly mature industrial applications provide favorable supports for photonic quantum technologies. Currently, fibre optical communication technology has become the pillar for information transmission, and photonic integration and optoelectronic integration for information processing has developed rapidly in recent years for the future all-optical network. Therefore, optical quantum systems are becoming a promising platform for quantum information processing \cite{O'brien2009,Flamini2018,Slussarenko2019}.

To enhance the complexity of quantum optical experiments, optical quantum systems tend to use photonic integrated circuits \cite{Wang2020}. Compared to systems that use discrete optical components on an optical table, the integrated photonic devices enable localize and manipulate photons
at micro/nano scales, and thus greatly improve the stability and scalability of quantum optical experiments and provide a complex and compact quantum photonics approach for quantum communication, sensing and computing applications \cite{Lu2021}. Therefore, the integrated techniques will lead quantum applications moving out of the laboratory and into large-scale and practical. Multiple optical waveguide materials, such as silica \cite{Politi2008,Jiang2017,Zhang20192,Li2022}, silicon (Si) \cite{Silverstone2016}, silicon nitride (SiN) \cite{Zhang2018,Lu2019,Taballione2021,Ren2022}, lithium niobate \cite{Jin2014,Wang20172,Xu2022}, and techniques, such as lithography \cite{Bojko2011}, laser-writing \cite{Sugioka2014}, have been developed. Among those materials, silicon photonics is a good candidate because of its easy preparation, high integration density and excellent optical properties \cite{Rudolph2017}. Also thanks to the leadership of classical silicon photonics in large-scale photonic integrated circuits, the performance of silicon-based devices has been rapidly improved \cite{Thomson2016}.

Currently, photonic devices on the silicon chip for various quantum applications have been developed, sucn as high-efficiency chip-fiber coupler \cite{Son2018}, large-scale programmable quantum photonic circuits \cite{Harris2016} and single-photon detectors \cite{Pernice2012}. However, scalable quantum  applications put forward new and higher requirements for integrated photonic devices. For example, high isolation on-chip filter is needed to separate pump and single-photon level signals, and modulator should also operate at cryogenic condition. In this paper, we review the relevant research results and state-of-the-art technologies on the silicon photonic chip for scalable quantum applications. Since silicon optical chips can be easy prepared, their applications are very wide and diverse. Here we just review common complementary metal oxide semiconductor (CMOS) compatible silicon-on-insulator (SOI) chips with telecom optical band O + C. We're glad to see that properties of some integrated components have already met the requirements for further quantum application and expansion. Furthermore, we point out the challenges ahead and further research directions. This review mainly focus on advances of silicon photonic devices, and for more information about related novel quantum experiments, we recommend some related good reviews, such as preparation of photon-pair sources with silicon photonics \cite{Feng2020,Chen2021}, advances of quantum experiments on the silicon chip \cite{Wang2020,Adcock2020}, and hybrid integrated techniques for quantum applications \cite{Elshaari2020,Kim2020}.

The review is structured as follows: Section II reviews the state-of-the-art technologies of basic components on the silicon chip, such as on-chip photon sources, detectors, chip interconnects and so on. In Section III, we introduce some key and fundamental quantum information processing in silicon photonic circuits and review research works on multiphoton and high dimensional applications, quantum error correction, quantum key distribution as well as quantum state teleportation between chips. These excellent works leads to scalable quantum computing and communication. Finally, in Section IV, we outlook the future of silicon photonics for scalable quantum information applications and provide challenges for further scaling.      

\section*{II. Scalable techniques on silicon photonic chips}

Scalable quantum information processing needs to improve the complexity on the single chip as much as possible. Different degrees of freedom of photons should achieve universal quantum operation, and quantum photonic source, detector, logic operation and other core functions should be improved to high quality and be integrated on the same chip. Considering the limited chip size and difficulty in hybrid integration with different materials, multifunction chip can also be achieved 
by high efficiency optical interconnection for quantum communication and distributed quantum computing and metrology. Accordingly, we divide the state-of-the-art techniques in silicon photonics into two aspects to review, single-chip and interconnection technologies. The contents of this section include single-photon sources, photon detection, wavelength and mode division multiplexing and cryogenic techniques on the single chip and chip interconnects techniques.

 

\subsection{A. Single-photon sources}

\begin{figure*}[t]
\centering
\includegraphics[width=14cm]{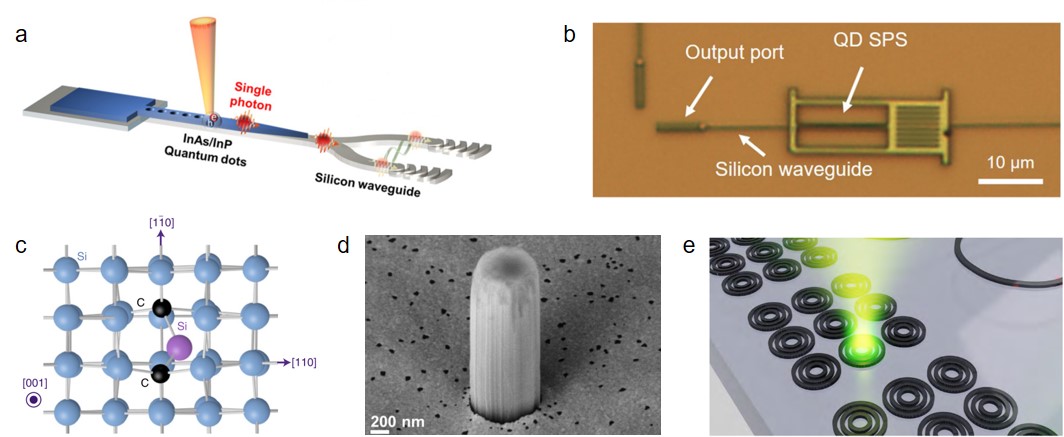}
\caption {\textbf{Solid-state quantum emitters in silicon photonics.} a. Position grown InAs/InP quantum dots on a silicon photonic chip by the pick-and-place technique. Adapted from Ref. \cite{Kim2017}. b. Integrate heterogeneous optical components with transfer-printing-based
approach. Adapted from Ref. \cite{Katsumi2019}. c. Atomic structure of the G-centre. Adapted from Ref. \cite{Redjem2020}. d. Si nanopillar including the G-centre. Adapted from Ref. \cite{Hollenbach2021}. e. Bullseye structures used to enhance vertical coupling of G-centers. Adapted from Ref. \cite{Prabhu2022}. } 
\label{fig1}
\end{figure*}

\begin{table*}
\caption{\textbf{State-of-the-art photonic sources at telecommunication wavelengths.} For herald single-photon sources, pump power normalized pair generation rate are given. While for solid-state emitters, saturated emission rate are given. PGR: pair generation rate; SER: saturated emission rate: CAR: coincidence-to-accident ratio. $\ast$ indicates the use of pulsed laser pump.}
\begin{tabular}{ |p{1cm}|p{3.2cm}|p{2cm}|p{2.5cm}|p{1cm}|p{1.5cm}|p{4cm}|  }
 \hline
 \makecell[c]{Ref.}& \makecell[c]{Structure}	&\makecell[c]{Bandwidth}&\makecell[c]{PGR/SER}&\makecell[c]{CAR}&\makecell[c]{$g^2(0)$}&\makecell[c]{Wavelength}\\
 \hline
   \makecell[c]{\cite{Li2017}}	&\makecell[c]{Single-mode waveguide}	& \makecell[c]{100$\,$GHz}&	\makecell[c]{0.7$\,$MHz.mW$^{-2}$}&
 \makecell[c]{80}&\makecell[c]{-- --}&\makecell[c]{1538.2$\,$\&$\,$1562.2$\,$nm}\\
 \hline
 \makecell[c]{\cite{Ma2017}}& \makecell[c]{Micro-ring resonator}	&\makecell[c]{2.1$\,$GHz}&	\makecell[c]{149$\,$MHz.mW$^{-2}$}&\makecell[c]{12105}&\makecell[c]{0.00533}&\makecell[c]{1535.5$\,$\&$\,$1574.7$\,$nm}	\\
  \hline
 \makecell[c]{\cite{Paesani2020} $\ast$}&	\makecell[c]{Multimode waveguide}	& \makecell[c]{4$\,$nm}	&\makecell[c]{18.6$\,$MHz.mW$^{-2}$}&\makecell[c]{-- --}&\makecell[c]{0.053}&\makecell[c]{1516$\,$\&$\,$1588$\,$nm}\\
  \hline
 \makecell[c]{\cite{Zhou2018}}	&\makecell[c]{GaN Defect}	& \makecell[c]{3--50$\,$nm}&	\makecell[c]{1.5$\,$MHz}&\makecell[c]{-- --}&\makecell[c]{0.05}&\makecell[c]{1085--1340$\,$nm}\\
  \hline
 \makecell[c]{\cite{Zhao2021} $\ast$}&	\makecell[c]{2D MoTe$_2$}	& \makecell[c]{8.5--37$\,$nm}&\makecell[c]{-- --}&\makecell[c]{-- --}& \makecell[c]{0.058}&\makecell[c]{1080--1550$\,$nm}\\
 \hline
 \makecell[c]{\cite{Hollenbach2020}}&	\makecell[c]{G center}	& \makecell[c]{0.5$\,$nm}	&\makecell[c]{99$\,$kHz}&\makecell[c]{-- --}&\makecell[c]{0.07}&\makecell[c]{1278$\,$nm}\\
\hline
 \makecell[c]{\cite{Higginbottom2022}}&	\makecell[c]{T center}	& \makecell[c]{255$\,$MHz}	&\makecell[c]{-- --}&\makecell[c]{-- --}&\makecell[c]{0.2}&\makecell[c]{1326$\,$nm}\\
\hline
 \end{tabular}
 \end{table*}

Scalable quantum photonic information processing requires multiple high-quality single photon sources. Today, there are two promising approaches to achieving a near-deterministic single photon source, one is based on multiplexing an array of probabilistic parametric photon pair sources, and the other is based on solid-state single-photon emitters.

With strong third-order nonlinear response, silicon waveguides can be directly used to prepare photon pair sources via spontaneous four-wave mixing process (SFWM) \cite{Li2017}. Many techniques were developed to improve photonic source quality, such as using micro-ring resonators \cite{Ma2017,Liu2020,Burridge2020} and introducing special phase-matching condition \cite{Paesani2020}. More details are given in Table I. Upon pair generation, one of the photons is detected to herald the presence of its partner. Although each photon pair source is probabilistic, multiple heralded single photons can be dynamically switched to one single output mode, and thus increasing the output probability. Any degree of freedom of the photon pairs can be used for multiplexing, such as path, frequency and time bins. Also, these can be combined to multiplicatively increase the number of multiplexed modes. In ref. \cite{Collins2013}, a 63.1\% efficiency increase in the heralded single-photon output has been demonstrated through multiplexing photons generated in two silicon waveguides. In ref. \cite{Zhang2015}, the enhancement was increased to 90$\pm$5\% by using a silicon waveguide pumped by time and wavelength division multiplexed pulses. In ref. \cite{Xiong2016}, 100\% enhancement was achieved by multiplexing photons from four temporal modes. Recently, although not within silicon photonics, a factor of 9.7 enhancement in efficiency has been realized using optical delay to multiplex 40 heralded single-photon sources \cite{Kaneda2019}. 

Despite these developments, on-chip integration of multiplexing systems is still an outstanding challenge, and the main obstacle is the device loss. For example, assuming that 1 ns delay line corresponds to 10 cm long silicon waveguide, and the waveguide has the state-of-the-art low loss of 0.08 dB/m \cite{Lee2012}, 100 ns delay line will introduce 0.8 dB loss, which is still much higher than that in commercial fiber. The integrated high-speed switch, one necessary component in multiplexing systems, usually needs ion doping or hybrid integration with other electro-optic materials such as lithium niobate \cite{He2019}, and these post-processing processes always introduce extra losses. Significant reduction in these losses will enable multiple near-deterministic single photon sources achieved on the silicon chip with probabilistic nonlinear parametric processes.

Another approach, the solid-state emitters generate deterministic single photons and can be integrated or transferred on the silicon photonic circuits. Since silicon is opaque below around 1100 nm, single-photon emitters integrated with silicon would need to radiate photons beyond this wavelength limit. Multiple quantum dots at telecommunication wavelengths have been demonstrated \cite{Haffouz2018,Kim2017,Muller2018,Katsumi2018,Katsumi2019,Zhou2018,Redjem2020,Zhao2021}. Among them, the most common are semiconductor quantum dots including InAsP \cite{Haffouz2018}, InAs/InP \cite{Kim2017,Muller2018} and InAs/GaAs \cite{Katsumi2018,Katsumi2019}, etc. By using emission-wavelength-optimized waveguides, ref. \cite{Haffouz2018} demonstrated photon emission from single InAsP quantum dot with large tuning range from 880 to 1550 nm. 
With hybrid integration techniques, transfer semiconductor quantum dots on the silicon waveguides (Fig. 1a and 1b) has been implemented with $g^2(0)$ values around 0.3 \cite{Kim2017,Katsumi2018,Katsumi2019}. By using localized defects in the gallium nitride crystal, high-quality solid-state quantum emitters in the telecom range has been achieved \cite{Zhou2018}. Even at room temperatures, $g^2(0) = 0.05$ was obtained with continuous wave (CW) laser excitation. Two-dimensional (2D) materials show many amazing properties and have also been used to produce quantum emitters, for example, telecom-wavelength single-photon emitters via coupling 2D molybdenum ditelluride (MoTe$_2$) to nano-pillar arrays have been reported with $g^2(0) = 0.058$ and $g^2(0) = 0.181$ under pulsed and CW laser excitation, respectively \cite{Zhao2021}. Colour centers, known as G centers and T centers, which originate from carbon-related defect in silicon, are becoming another kind of candidates with telecom-O band radiation \cite{Redjem2020,Hollenbach2020,Bergeron2020,Higginbottom2022}. Different from other methods, 
colour centers can be directly integrated into silicon waveguides without hybrid integration for large-scale quantum photonic information applications \cite{Hollenbach2021,Prabhu2022}. For example, tens of thousands of individually addressed photon-spin qubits have been demonstrated with T centers, which will provide photonic links between spin qubits and greatly advance quantum information networks \cite{Higginbottom2022}.

To integrate solid-state emitters with photonic waveguides, many approaches are proposed such as nanoscale positioning approaches \cite{Liu20212}. For the important coupling efficiency,
ref. \cite{Katsumi2018} showed total single-photon coupling efficiency of 63\% experimentally and in theory higher than 99\%. 
In ref. \cite{Prabhu2022}, emitters were directly integrated in silicon waveguides and showed 40\% coupling efficiency in simulation. In addition, the use of optical micro-cavities deserves further consideration to improve the extraction efficiency \cite{Peng2017,Liu20213,Wei2022}. To further scaling up, simultaneous operation of multiple emitters is necessary. However, no two solid-state quantum emitters are alike when being produced experimentally. To keep coherence among them, wavelength tunability is necessary. Electronic integration or induced strain have been utilised to tune the wavelength and quantum interference between photons generated from independent quantum emitters have been demonstrated \cite{Patel2010,Flagg2010}. In nanophotonic devices, the wavelength tuning of the selected integrated quantum emitter have been realized \cite{Elshaari2018,Machielse2019,Katsumi2020}. The rapid development of processing and tuning techniques has enhanced the indistinguishability of photons between different emitters and increased the number of emitters on-chip. Though not in the telecom band, quantum interference of photons from two remote quantum dots with a visibility of 93.0\% \cite{Zhai2022} and large-scale integration of artificial atoms in hybrid AlN photonic circuits \cite{Wan2020} have been demonstrated recently.

\subsection{B. Photon Detection}

\begin{figure*}[t]
\centering
\includegraphics[width=15cm]{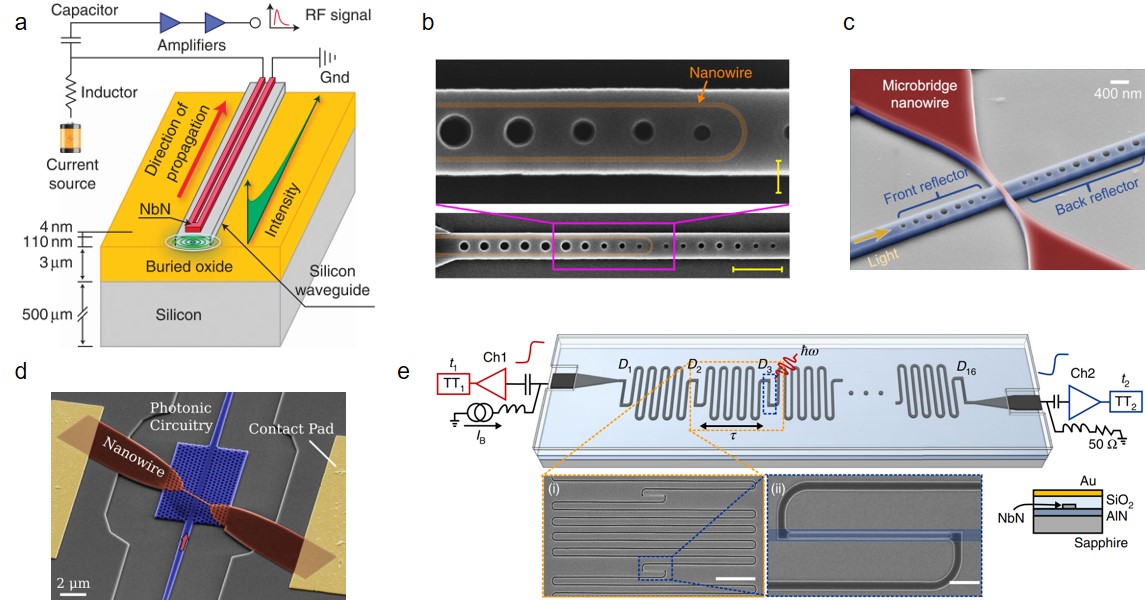}
\caption {\textbf{Integrated SNSPDs for single-photon detection.} a. Principle of travelling wave coupling. Adapted from Ref. \cite{Pernice2012}. b. SNSPD within a high-quality factor microcavity. Adapted from Ref. \cite{Akhlaghi2015}. c. Cavity-integrated SNSPD. Adapted from Ref. \cite{Vetter2016}. d. SNSPD implemented in a two-dimensional photonic crystal cavity. Adapted from Ref. \cite{Munzberg2018}. e. A typical chain of single-photon detector segments for signal multiplexing and number resolution. Adapted from Ref. \cite{Zhu2018}. } 
\label{fig2}
\end{figure*}

\begin{table*}
\caption{\textbf{State-of-the-art techniques for integrated single-photon detection.} Except ref. \cite{Llin2020}, whose operating wavelength is 1310$\,$nm, all other operating wavelengths are around 1550$\,$nm. $\ast$ represents the system detection efficiency.}
\begin{tabular}{ |p{1cm}|p{2.8cm}|p{1.2cm}|p{2.5cm}|p{1.5cm}|p{2cm}|p{2.5cm}|   }
 \hline
 \makecell[c]{Ref.}& \makecell[c]{Detection Efficiency}	&\makecell[c]{Jitter}&\makecell[c]{Dark Count Rate}&\makecell[c]{Reset Time}&\makecell[c]{Temperature}&\makecell[c]{Number Resolving}\\
 \hline
 \makecell[c]{\cite{Pernice2012}}& \makecell[c]{91\%}	&\makecell[c]{18$\,$ps}&	\makecell[c]{50$\,$Hz}&\makecell[c]{505$\,$ps}&\makecell[c]{1.7$\,$K}&\makecell[c]{-- --}	\\
  \hline
 \makecell[c]{\cite{Akhlaghi2015}}&	\makecell[c]{100\%}	& \makecell[c]{55$\,$ps}	&\makecell[c]{0.1$\,$Hz}&\makecell[c]{7$\,$ns}&\makecell[c]{2.05$\,$K}&\makecell[c]{-- --}\\
  \hline
 \makecell[c]{\cite{Vetter2016}}	&\makecell[c]{30\%}	& \makecell[c]{32$\,$ps}&\makecell[c]{1$\,$Hz}&\makecell[c]{510$\,$ps}&
 \makecell[c]{1.7$\,$K}&\makecell[c]{-- --}\\
 \hline
 \makecell[c]{\cite{Munzberg2018}}	&\makecell[c]{70\%}	& \makecell[c]{480$\,$ps}&	\makecell[c]{0.1$\,$mHz}&\makecell[c]{480$\,$ps}&\makecell[c]{1.6$\,$K}&\makecell[c]{-- --}\\
  \hline
 \makecell[c]{\cite{Zhu2020}}&	\makecell[c]{5.6\% $\ast$}	& \makecell[c]{16.1$\,$ps}&\makecell[c]{2$\,$Hz}&\makecell[c]{85.8$\,$ns}&\makecell[c]{1.0$\,$K}& \makecell[c]{4}\\
 \hline
 \makecell[c]{\cite{Llin2020}}&	\makecell[c]{29.4\%}	& \makecell[c]{134$\,$ps}	&\makecell[c]{100$\,$kHz}&\makecell[c]{-- --}&\makecell[c]{125$\,$K}&\makecell[c]{-- --}\\
\hline
\makecell[c]{\cite{Fang2020}}&	\makecell[c]{60.1\%}	& \makecell[c]{-- --}	&\makecell[c]{340$\,$kHz}&\makecell[c]{88$\,$ns}&\makecell[c]{300$\,$K}&\makecell[c]{-- --}\\
\hline
 \end{tabular}
 \end{table*}
 
Photon detector converts optical signals into electrical signals and is an essential part for photonic quantum information applications. For squeezed light measurement, germanium (Ge) photodetectors integrated on silicon were developed \cite{Benedikovic2019} and homodyne detectors have been realized based on them \cite{Rafaelli2018}. Furthermore, by interfacing silicon photonics with integrated amplification electronics, complete integrated homodyne detectors have been demonstrated \cite{Tasker2021} and a shot-noise-limited bandwidth of more than 20 GHz has been achieved \cite{Bruynsteen2021}.

For single-photon level detection, superconducting nanowire single-photon detectors (SNSPDs) show excellent performance, such as near unity system detection efficiency \cite{Reddy2020} GHz maximum count rate \cite{Pernice2012} and picosecond level temporal resolution \cite{Korzh2020}. SNSPD uses superconducting material that works below its critical temperature. Even if only one photon hits, the energy is enough to excite the material, and generate voltage pulses subsequently. A range of superconducting materials have been developed, including NbN \cite{Pernice2012}, NbTiN \cite{Akhlaghi2015}, MoSi \cite{Li2016} and WSi \cite{Buckley2017}, etc. In vertical optical coupling between fibers and SNSPDs, fiber end faces are parallel to the SNSPDs' photonsensitive surfaces, and photons are vertically incident to the nanowires. By optimizing the device's vertical optical stack design, as well as the coupling of the guided fiber mode to the active detection area of the device, 98\% system detection efficiency has been achieved \cite{Reddy2020}. For coupling between waveguides and SNSPDs, travelling wave coupling is used, as shown in Fig. 2a, which can achieve efficient detection of photons in the optical waveguide. This hybrid integration method directly places SNSPDs into silicon photonics, thus greatly enhancing the scalability of quantum photonic integrated circuits. For example, in ref. \cite{Pernice2012}, ballistic photon transport in silicon ring resonators has been demonstrated by exploiting high temporal resolution detectors. Besides, SNSPDs are integrated with high-quality factor microcavities to increase the coupling efficiency (Fig. 2b) \cite{Akhlaghi2015} and decrease the dark count rate (Fig. 2c and 2d) \cite{Vetter2016,Munzberg2018}.

Typically, single-photon detectors operate in Geiger mode, that is, the electrical signal generated by the read-out circuit indicating the detection of one photon. However, in many experiments, it is desirable to use a detector that can resolve the number of photons. By analyzing the relationship between photon numbers and resistance of the nanowire, multi-photon detection was achieved with one convertioanl SNSPD \cite{Cahall2017}. SNSPDs with photon number resolving ability has also been achieved on the AlN platform through exploring the pulse shapes of the detector output (Fig. 2e) \cite{Zhu2018}. Besides, with an integrated impedance-matching taper on the superconducting nanowire, the detector’s output amplitude turned into more sensitive to the number of photons, and detector with photon number resolving up to four, 16.1 ps timing jitter and \textless$\,$2$\,$Hz dark count rate has been demonstrated \cite{Zhu2020}.

The high-quality of SNSPDs boosts various novel applications, such as quantum computational advantage using photons\cite{Zhong2020}, imaging and spectroscopy \cite{Zhao2017,Wollman2019,Cheng2019}, and integrated quantum key distribution \cite{Beutel2021,Zheng2021}. 
Despite great progress achieved, SNSPDs cannot work without bulky and expensive cryogenic systems (\textless$\,$4$\,$K). Other types, such as germanium-on-silicon \cite{Vines2019,Llin2020} and InGaAs/InP single-photon detectors \cite{Zhangzhang2015,Fang2020}, are potential alternatives. For detection efficiency, germanium-on-silicon detectors have recently been demonstrated to be 38\% at 1310$\,$nm at 125$\,$K \cite{Vines2019}, and InGaAs/InP detectors to be 60.1\% at 1550$\,$nm at 300$\,$K. Further reduction of dark count rate will broaden their appeal in many quantum applications. More details about integrated single-photon detectors are summarized in Table II. In addition to these conventional devices, single-photon detection based on low-dimensional materials is emerging and has shown superior performance \cite{Wang2022}. 

\subsection{C. Wavelength and mode division multiplexing}


\begin{figure}[t]
\centering
\includegraphics[width=8cm]{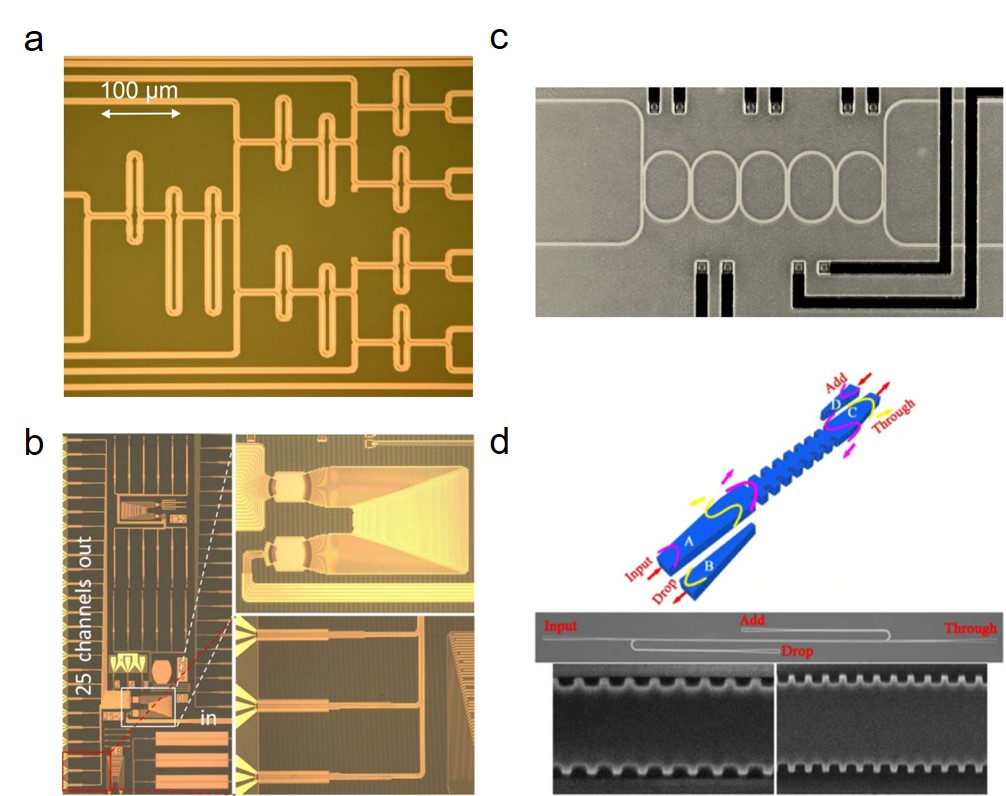}
\caption {\textbf{Wavelength division multiplexing techniques in silicon photonics.} a. Cascaded Mach-Zehnder demultiplexer. Adapted from Ref. \cite{Horst2013}. b. The wavelength division multiplexing receiver chip with an integrated arrayed waveguide grating. Adapted from Ref. \cite{Liuliu2019}. c. Coupled 5-ring silicon filter. Adapted from Ref. \cite{Ong2013}. d. Waveguide Bragg grating add-drop filter. Adapted from Ref. \cite{Qiu2017}.} 
\label{fig3}
\end{figure}

\begin{table*}
\caption{\textbf{State-of-the-art techniques for filting in silicon photonics.} MZI: Mach-Zehnder interferometer; WBG: waveguide Bragg grating; IL: insertion loss; FSR: free spectral range.}
\begin{tabular}{ |p{1cm}|p{3cm}|p{1.5cm}|p{1.5cm}|p{1cm}|p{1cm}|p{4cm}|  }
 \hline
 \makecell[c]{Ref.}& \makecell[c]{Structure}	&\makecell[c]{Size}&\makecell[c]{Contrast}&\makecell[c]{IL}&\makecell[c]{FSR}&\makecell[c]{Bandwidth}\\
 \hline
 \makecell[c]{\cite{Xia2007}}& \makecell[c]{High-order micro-ring}	&\makecell[c]{700$\,\upmu$m$^2$}&	\makecell[c]{40$\,$dB}&\makecell[c]{1.8$\,$dB}&\makecell[c]{18$\,$nm}&\makecell[c]{310$\,$GHz (1$\,$dB)}	\\
  \hline
 \makecell[c]{\cite{Ong2013}}&	\makecell[c]{High-order micro-ring}	& \makecell[c]{3000$\,\upmu$m$^2$}	&\makecell[c]{50$\,$dB}&\makecell[c]{3$\,$dB}&\makecell[c]{7.3$\,$nm}&\makecell[c]{11.6--125$\,$GHz (3$\,$dB)}\\
  \hline
 \makecell[c]{\cite{Liao2014}}	&\makecell[c]{Unbalanced MZIs}	& \makecell[c]{2$\,$mm$^2$}&	\makecell[c]{15$\,$dB}&
 \makecell[c]{9$\,$dB}&\makecell[c]{0.8$\,$nm}&\makecell[c]{0.61, 0.34 and 0.21$\,$nm (3$\,$dB)}\\
 \hline
 \makecell[c]{\cite{Qiu2017}}	&\makecell[c]{WBG}	& \makecell[c]{600$\,\upmu$m$^2$}&	\makecell[c]{35$\,$dB}&\makecell[c]{0.6$\,$dB}&\makecell[c]{-- --}&\makecell[c]{3$\,$nm (3$\,$dB)}\\
  \hline
 \makecell[c]{\cite{Harris2014}}&	\makecell[c]{Cascaded WBG}	& \makecell[c]{1280$\,\upmu$m$^2$}&\makecell[c]{65$\,$dB}&\makecell[c]{3$\,$dB}& \makecell[c]{-- --}&\makecell[c]{1--2$\,$nm (3$\,$dB)}\\
 \hline
 \makecell[c]{\cite{Oser2020}}&	\makecell[c]{Cascaded WBG}	& \makecell[c]{3105$\,\upmu$m$^2$}	&\makecell[c]{60$\,$dB}&\makecell[c]{2$\,$dB}&\makecell[c]{-- --}&\makecell[c]{5.5$\,$nm (3$\,$dB)}\\
\hline
 \end{tabular}
 \end{table*}

Due to the significant enhancement in demand for information processing and communication capability, much effort has been done to encode more information on one single chip. One way is to increase the integration density. Devices are developed to be more compact and their spacing to be narrower. For example, by using the concept of symmetry breaking, high-density waveguide superlattices with low crosstalk were demonstrated \cite{Song2015}.This concept has also been used to construct two-qubit quantum logic gate, and greatly reduced the footprint \cite{Zhang2021}. Recently, an ultradensely integrated multidimensional optical system with a footprint of 20$\,\times\,$30$\,\upmu$m$^2$ has been demonstrated based on inverse-designed dielectric metasurface network \cite{Zhou2022}. Another way is to introduce new degrees of freedom through multiplexing, including wavelength division multiplexing (WDM) and mode division multiplexing (MDM).

WDM is a technique that encodes information with different optical wavelengths as different channels. Structures like ring resonators, unbalanced MZIs, waveguide Bragg gratings (WBGs) and arrayed waveguide gratings (AWGs) are used in the WDM systems for the purpose of wavelength multiplexing and de-multiplexing \cite{Horst2013,Cheung2014,Tan2014,Chen2015,Munk2019,Liuliu2019}. Among them, 512 channel dense WDM \cite{Cheung2014}, channel spacing down to 17$\,$GHz \cite{Munk2019}, hybrid integration with detectors (Fig. 3b) \cite{Liuliu2019} have been demonstrated. Although not with silicon photonics, high-speed quantum key distribution with WDM on integrated photonic circuits has been reported \cite{Price2018}. 

\begin{figure*}[t]
\centering
\includegraphics[width=16cm]{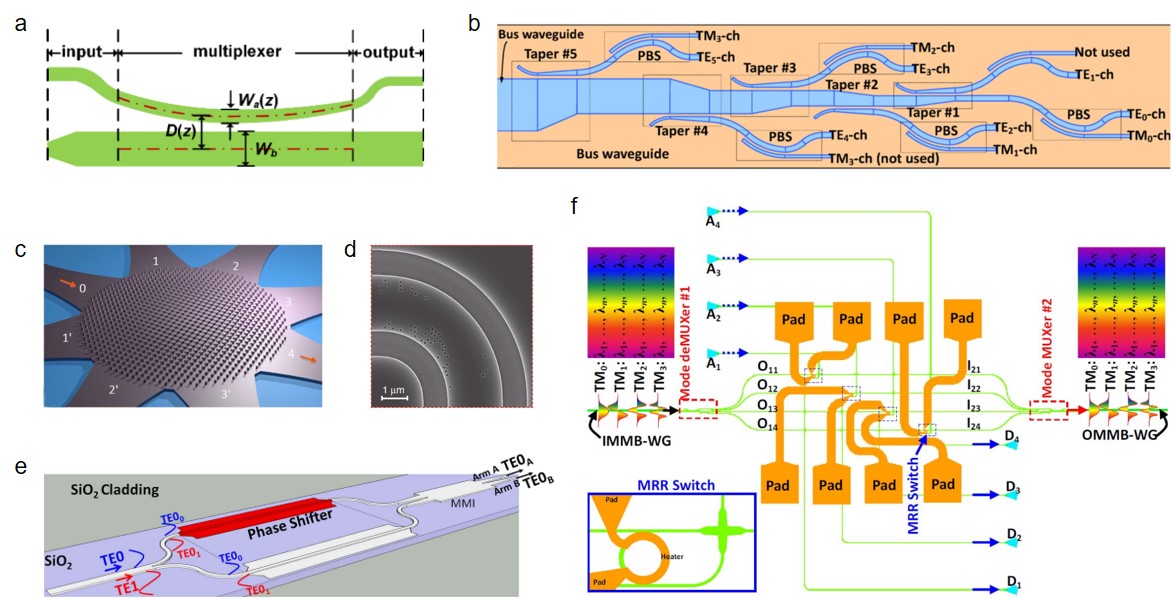}
\caption {\textbf{Mode division multiplexing techniques in silicon photonics.} a. Mode (de)multiplexer with adiabatic taper. Adapted from Ref. \cite{Guo2017}. b. 10-channel mode (de)multiplexer. Adapted from Ref. \cite{Dai2018}. c. A multiport multimode waveguide crossing using a metamaterial-based Maxwell's fisheye lens. Adapted from Ref. \cite{Xu2018}. d. The digital meta-structure-based multimode bending. Adapted from Ref. \cite{Liuy2019}. e. High-speed optical two-mode switch. Adapted from Ref. \cite{Xiong2017}. f. Reconfigurable optical add-drop multiplexer for hybrid wavelength/mode-division-multiplexing systems. Adapted from Ref. \cite{Wang2017}.} 
\label{fig4}
\end{figure*}

In addition to increasing channel capacity, WDM technique is also used as filters for the spectrum manipulation in various optical systems \cite{Liu2021}. With unbalanced MZIs, an programmable filter has been demonstrated and showed the tunability of the filter central wavelength, bandwidth and variable passband shape \cite{Liao2014}. With high-order ring resonators, on-chip filters are becoming ultra-compact, ultra-high-contrast and show high flexibility \cite{Xia2007,Ong2013,Chen2014,Liu2019}. For example, 5$^{\rm{th}}$ order ring resonator optical filters showed out-of-band rejection ratio of 40$\,$dB, and insertion loss of only 1.8$\,$dB within a footprint of 700$\,\upmu$m$^2$ (Fig. 3c) \cite{Xia2007}. To increase the free spectral range (FSR), a resonator with a small radius of 0.8$\,\upmu$m was demonstrated and a record large FSR of 93$\,$nm was achieved recently \cite{Liu2019}. WBGs are FSR free and could also be employed to build filters. The add-drop structure showed high-contrast and the filter bandwidths changed with the waveguide widths (Fig. 3d) \cite{Qiu2017}.  More details about filting in silicon photonics are summarized in Table III. For scalable quantum information applications, these filters will be employed to filter out the pump light before detection or separate single photons at different wavelengths, as reported in refs. \cite{Harris2014,Oser2020}. 

MDM is an emerging technique that uses the high-order transverse waveguide modes of multimode waveguides to encode more information \cite{Li2019}. Since multimode waveguides still support multiple wavelengths, this technique is compatible with WDM to further increase the channel capacity \cite{Luo2014,Dai2015,Wang2017}. As shown in Fig. 4, various multimode silicon photonic devices have been developed for low-loss and low-crosstalk light manipulation with multiple mode-channels in MDM systems, such as mode (de)multiplexer \cite{Dai2013,Guo2017}, grating couplers \cite{Lai2018}, high-speed switches \cite{Xiong2017}, sharp waveguide bends and mode-independent crossings \cite{Li2018,Xu2018,Liuy2019}. At present, mode (de)multiplexer including six mode-channels of TE polarization and four mode-channels of TM polarization has been realized \cite{Dai2018}. In addition, by using multimode optical waveguides with higher-order modes, some special silicon photonic devices that cannot be realized only with fundamental modes have been demonstrated, such as add-drop optical filters based on multimode Bragg gratings \cite{Qiu2017,Oser2020}. 

The utilization of higher-order modes for quantum photonics has stimulated many novel applications. For example, in ref. \cite{Feng2016}, the transverse waveguide-mode degree of freedom was introduced for quantum encoding, and on-chip coherent conversion of photonic entangled quantum states between path, polarization, and transverse waveguide-mode degrees of freedom was demonstrated. In ref. \cite{Mohanty2017}, quantum interference between the guided modes was demonstrated within a multimode optical waveguide. Recently, with the help of two newly developed multimode devices, transverse mode-encoded 2-qubit logic quantum gate was realized \cite{Feng2022}, and showed the potential for universal transverse mode-encoded quantum operations and large-scale multimode multi-degrees of freedom quantum systems. The special inter-modal phase matching conditions were used to prepare quantum photonic sources. In particular, an on-chip transverse-mode entangled photon pair source via SFWM in a multimode waveguide was demonstrated \cite{Feng2019}. Furthermore, by exploiting special excitation scheme in low-loss multimode waveguides, quantum photonic source was engineered to high spectral purity and mutual indistinguishability through inter-modal SFWM \cite{Paesani2020}.

\subsection{D. Cryogenic techniques}

\begin{figure}[t]
\centering
\includegraphics[width=7cm]{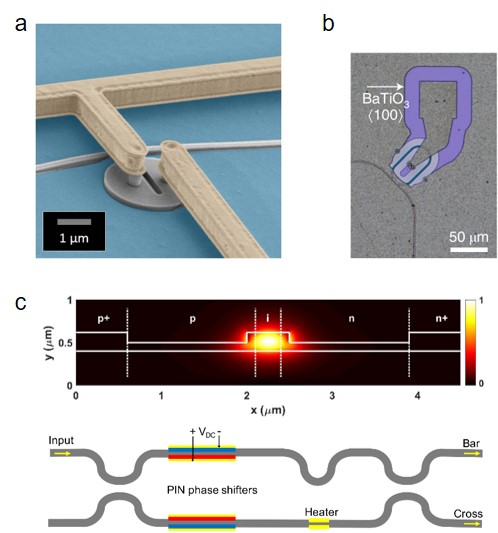}
\caption {\textbf{Silicon photonic modulators at cryogenic temperatures.} a. The plasma dispersion micro-disk modulator. Adapted from Ref. \cite{Gehl2017}. b. The BaTiO$_3$–Si racetrack resonator. Adapted from Ref. \cite{Eltes2020}. c. The integrated PIN junction modulator and unbalanced Mach–Zehnder interferometer composed of the modulator. Adapted from Ref. \cite{Chakraborty2020}.} 
\label{fig5}
\end{figure}

The large-scale expansion of quantum photonic integrated circuits requires the integration of all functions on one single chip including quantum photonic sources, state manipulation and photon detection. Because common and efficient SNSPDs must work at cryogenic temperatures, for example, 2$\,$K, quantum photonic sources and state manipulation processes should also extend to the same temperature condition. Besides, in quantum networks, optical interfaces with other quantum systems such as semiconductor and superconducting quantum computing systems also need to operate the whole system at cryogenic temperatures.

While passive optical components can often be used directly at low temperatures, implementation of active components and nonlinear processes has been a prominent challenge. At present, quantum state manipulation structures used in quantum photonic chips at ambient conditions are generally based on thermo-optic effect. However, the thermo-optic coefficient of silicon decreases significantly
at low temperatures \cite{Komma2012}. In particular, when the temperature is several kelvin, the thermo-optic coefficient is four orders of magnitude lower than that at room temperatures, which makes thermo-optic modulators difficult to work. Another problem is that as temperatures drop, so does the cooling capacity available of the cooling system. 
Therefore, modulators operating at cryogenic conditions should maintain very low power consumption and heat production. 

To date, some works have been reported on cryogenic modulation \cite{Gehl2017,Eltes2020,Chakraborty2020} with silicon photonics. In ref. \cite{Gehl2017}, by doping at higher concentrations, one microdisk modulator (Fig. 5a) prepared by plasma dispersion effect has been achieved at temperature 4.8 K and transmitting data rates up to 10 Gb/s has been achieved. As the first success implementation, this work opens the door for using silicon photonics to interface other systems at cryogenic conditions. Besides, electro-optic modulation at temperature 4 K by using the Pockels effect of integrated barium titanate (BaTiO$_3$) devices has been reported (Fig. 5b)) \cite{Eltes2020}. This material showed an effective Pockels coefficient of 200 pm V$^{-1}$ at 4 K, and the fabricated devices showed high electro-optic bandwidth (30 GHz), ultralow-power consumption and high-speed data modulation (20 Gb/s). 
Another implementation employed DC Kerr effect of the silicon waveguides and achieved phase modulation at a temperature of 5 K at GHz speeds (Fig. 5c) \cite{Chakraborty2020}. Despite these progresses, these modulation devices are still in the initial research stage, and the relevant applications in quantum information research have not been demonstrated. For large-scale quantum information applications, they should be more compact and have lower excess loss. More recently, programmable Mach–Zehnder meshes have been realized by using aluminium nitride (AlN) piezo-optomechanical actuators coupled to SiN waveguides \cite{Dong2022}, which can also operate at cryogenic conditions. Though the modulators operate in the visible band and on the SiN platform, the work demonstrates the possibility of on-chip large-scale cryogenic modulation.

In addition to cryogenic modulation, some works have also been reported on cryogenic nonlinear processes in silicon photonics \cite{Pernice2011,Sun2013,Sinclair2019}. In ref. \cite{Sinclair2019}, the temperature dependence of the two-photon absorption and optical Kerr nonlinearity of a silicon waveguide with temperatures from 5.5 to 300 K was measured, and the nonlinear figure of merit was found to be improved at cryogenic temperatures. Quantum applications, however, such as quantum photonic sources preparation by cryogenic SFWM, has not been demonstrated yet.

\subsection{E. Chip interconnects}
Chip interconnection plays the key role in building large-scale quantum networks. In photonic quantum technologies, chip interconnection needs efficient transfer of optical signals among different optical components. However, due to the mismatch in the effective mode sizes between different components such as fibers and silicon waveguides, special coupling structures are needed \cite{Son2018}. 

\begin{figure*}[t]
\centering
\includegraphics[width=16cm]{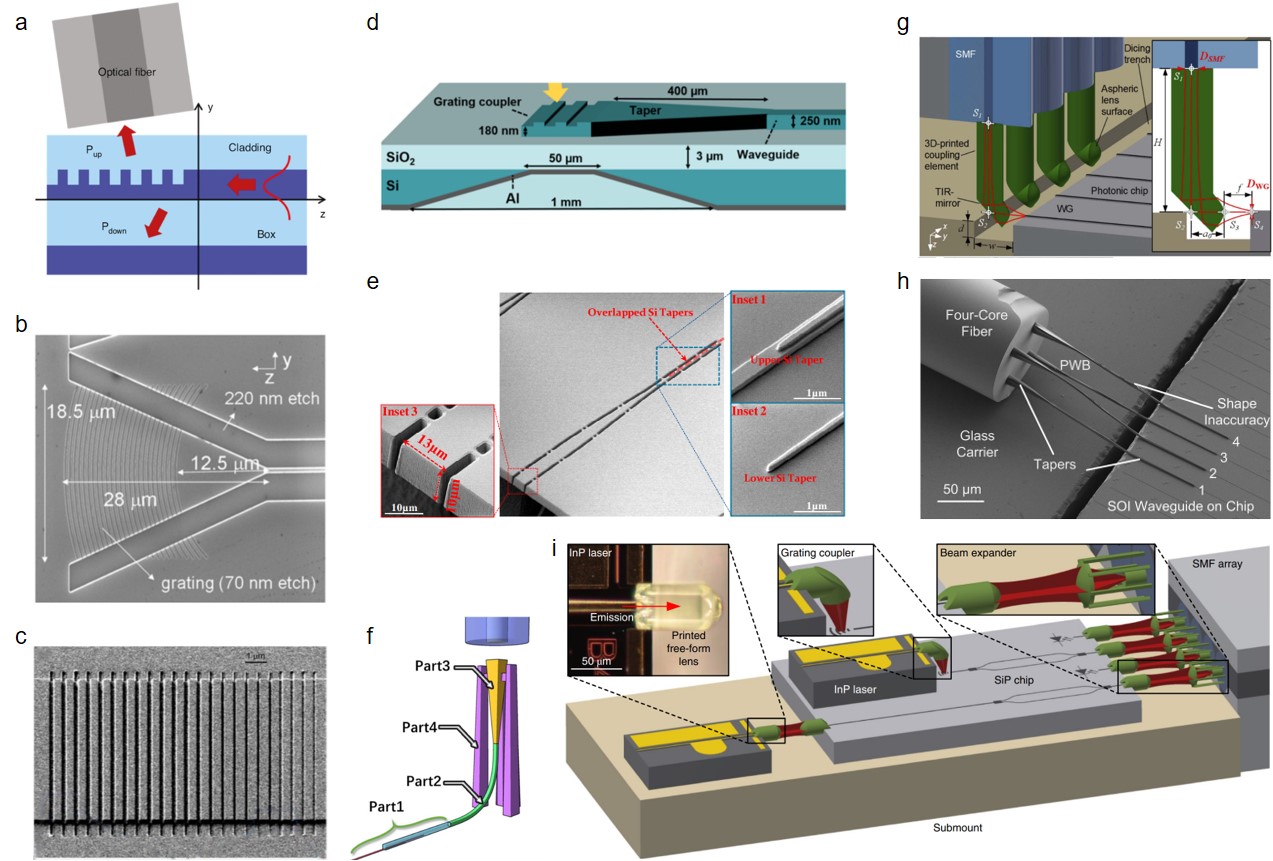}
\caption {\textbf{Chip interconnection techniques in silicon photonics.} a. Diffraction grating-based coupling structure. Adapted from Ref. \cite{Son2018}. b. The focusing grating. Adapted from Ref. \cite{Van2007}. c. The double-etched apodized waveguide grating coupler. Adapted from Ref. \cite{Li2013}. d. The grating coupler with a single aluminum backside mirror. Adapted from Ref. \cite{Hoppe2020}. e. The mode-size converter as end coupler. Adapted from Ref. \cite{Fang2011}. f. 3D coupler structure. Adapted from Ref. \cite{Luo2020}. g. The 3D-printed optical probes on the fiber end faces. Adapted from Ref. \cite{Trappen2020}. h. Fiber cores and different silicon waveguides connected by photonic wire bonds. Adapted from Ref. \cite{Lindenmann2015}. i. In situ 3D nanoprinted free-form lenses and expanders. Adapted from Ref. \cite{Dietrich2018}.} 
\label{fig6}
\end{figure*}

\begin{table*}
\caption{\textbf{State-of-the-art techniques for chip interconnection in silicon photonics.} For wire bonding and in situ 3D nanoprinting techniques, different coupling efficiencies were reported for coupling different structures. Here, the maximum values are given.}
\begin{tabular}{ |p{1.5cm}|p{4cm}|p{4cm}|p{4.5cm}|  }
 \hline
 \makecell[c]{Ref.}&\makecell[c]{Technique}&\makecell[c]{Loss}&\makecell[c]{Bandwidth}\\
 \hline
 \makecell[c]{\cite{Ding2014}}&\makecell[c]{Grating coupling}&\makecell[c]{0.58$\,$dB (TE)}&\makecell[c]{71$\,$nm (3$\,$dB)}\\
 \hline
  \makecell[c]{\cite{Pu2010}}&\makecell[c]{End coupling (tapered fiber)}&\makecell[c]{0.36$\,$dB (TM); 0.66$\,$dB (TE)}&\makecell[c]{\textgreater 80$\,$nm (1$\,$dB)}\\
 \hline
   \makecell[c]{\cite{Fang2011}}	&\makecell[c]{End coupling (SMF)}&\makecell[c]{2.0$\,$dB (TM); 1.2$\,$dB (TE)}&\makecell[c]{\textgreater 120$\,$nm  (1$\,$dB)}\\
 \hline
    \makecell[c]{\cite{Jia2018}}	&\makecell[c]{End coupling (SMF)}&\makecell[c]{1.3$\,$dB (TM); 0.95$\,$dB (TE)}&\makecell[c]{\textgreater 100$\,$nm  (1$\,$dB)}\\
 \hline
     \makecell[c]{\cite{Luo2020}}	&\makecell[c]{3D vertical coupling}&\makecell[c]{1$\,$dB (TE and TM)}&\makecell[c]{170$\,$nm (TE); 104$\,$nm (TM)  (1$\,$dB)}\\
 \hline
      \makecell[c]{\cite{Trappen2020}}	&\makecell[c]{3D-printing}&\makecell[c]{1.9$\,$dB} &\makecell[c]{-- --}\\
 \hline
       \makecell[c]{\cite{Lindenmann2012,Lindenmann2015,Billah2018}}	&\makecell[c]{Wire bonding}&\makecell[c]{0.4$\,$dB} &\makecell[c]{-- --}\\
 \hline
         \makecell[c]{\cite{Dietrich2018}}	&\makecell[c]{In situ 3D nanoprinting}&\makecell[c]{0.6$\,$dB} &\makecell[c]{-- --}\\
 \hline
 \end{tabular}
 \end{table*}

Grating coupler is one common such structure, which enlarges the mode size of the waveguide by etching diffraction gratings on the waveguide surface \cite{Taillaert2002}. The cross-section view of the uniform surface-corrugated grating structure is shown in Fig. 6a, and the picture of one fabricated grating coupler is shown in Fig. 6b. Typically, the grating couplers have an alignment tolerance value of $\pm$2$\,\upmu$m for 1$\,$dB excess losses and over 20$\,$nm 1$\,$dB bandwidth. The relaxed spatial alignment tolerance facilitates multi-path alignment, which is one key part for scalable photonic information applications. Since quantum signals are difficult to be amplified, the losses should be reduced as much as possible. For the uniform grating structure, the grating couplers have a typical coupling efficiency of 5$\,$dB \cite{Van2007}. To improve the coupling efficiency, different principles and techniques were introduced. One method is to spatially vary fill factors and etch depths (Fig. 6c), which can achieve larger overlap integral values between the field profiles of light from the grating coupler structure and the optical fiber \cite{Tang2010,Li2013,Benedikovic2014,Li2014,Marchetti2017}. These apodized grating couplers achieved a typical coupling efficiency of 1$\,$dB. Considering that a substantial portion of the optical power input is lost in the substrate due to the diffraction downward from the grating structure (Fig. 6a), another rule is to improve the radiation directionality. Many techniques were developed, such as using overlay grating elements \cite{Vermeulen2010} and substrate metal mirrors (Fig. 6d) \cite{Hoppe2020}. Especially, one can improve both the overlap integral and directionality \cite{Taillaert2004}, and below 1$\,$dB coupling loss was achieved in this way \cite{Ding2014}. It is worth mentioning that such high-efficiency grating couplers have been employed and played an important role in multi-photon quantum information processing on the silicon photonic circuits \cite{Paesani2019,Llewellyn2020,Vigliar2021}. The grating couplers are typically polarization dependent. In order to utilize polarization degree of freedom to encode quantum information, 2D grating couplers are alway used, which are able to couple orthogonal polarization of light into separate waveguides \cite{Xue2019}.  

End coupler is another common structure for chip-fiber coupling. Designs of the tapered waveguide have been introduced to enlarge the effective mode size of the integrated silicon waveguides (Fig. 6e). Similar to grating couplers, the end couplers have also achieved below 1$\,$dB coupling losses with commercial single mode fibers \cite{Pu2010,Fang2011,Jia2018}. In addition, the end couplers can achieve greater 1$\,$dB bandwidth and support the simultaneous and efficient coupling of two polarization modes (Table IV). The locating on the chip edges facilitates direct packaging with the fiber array, despite requiring additional dedicated fabrication steps, such as chip dicing and polishing. Multi-channel packaging with end coupling has been demonstrated for deep learning \cite{Shen2017} and quantum transport simulations \cite{Harris2017}. Similar to end coupling, a three-dimensional (3D) fabricated polymer coupler showing coupling loss of 1$\,$dB was demonstrated  recently (Fig. 6f). \cite{Luo2020}. 

In addition to fabricating complex coupling structures on chip, many other efficient approaches are also worth considering. For example, 3D-printed optical probes on the fiber end faces (Fig. 6g), which can realize the detection of vertical cutting edge devices, were recently demonstrated with 1.9$\,$dB coupling loss \cite{Trappen2020}. Photonic wire bonding and in situ 3D nanoprinting are other novel and promising techniques for chip-scale multi-platform interconnects (Fig. 6h and 6i). Using polymer waveguides with 3D geometry, photonic wire bonding can bridge photonic circuits on different chips \cite{Lindenmann2012,Lindenmann2015,Billah2018,Rhee2022}. With this technique, 1.6$\,$dB coupling loss between silicon chips \cite{Lindenmann2012}, 1.7$\,$dB coupling loss between a four-core fiber and a silicon chip \cite{Lindenmann2015} and 0.4$\,$dB coupling loss between an indium phosphide chip and a silicon photonic chip \cite{Billah2018} were demonstrated. In situ 3D nanoprinting technique exploits direct-write two-photon laser lithography to create ultracompact elements such as lenses and expanders, which can be directly integrated onto surfaces of optical integrated devices \cite{Dietrich2018}. These elements can be optimized according to the coupling objectives, thus increasing the flexibility. In ref. \cite{Dietrich2018}, coupling efficiencies of up to 0.6$\,$dB between edge-emitting lasers and single-mode fibres and 2.5$\,$dB between lasers and passive chips were achieved. Introducing these two new technologies into integrated quantum photonics will greatly improve the chip complexity and integration.

\section*{III. Scalable quantum information applications}

After decades of development, silicon photonics has made a series of significant advances in quantum computing, quantum simulation, quantum communication and metrology, etc. Here we give a review of some important advances in recent years, and they are fundamental and have irreplaceable values in the extended applications of quantum information processing, including multiphoton and high-dimensional applications and quantum error correction on one single chip, and quantum key distribution and state teleportation among chips.

\subsection{A. Multiphoton and high-dimensional applications}

\begin{figure*}[t]
\centering
\includegraphics[width=16cm]{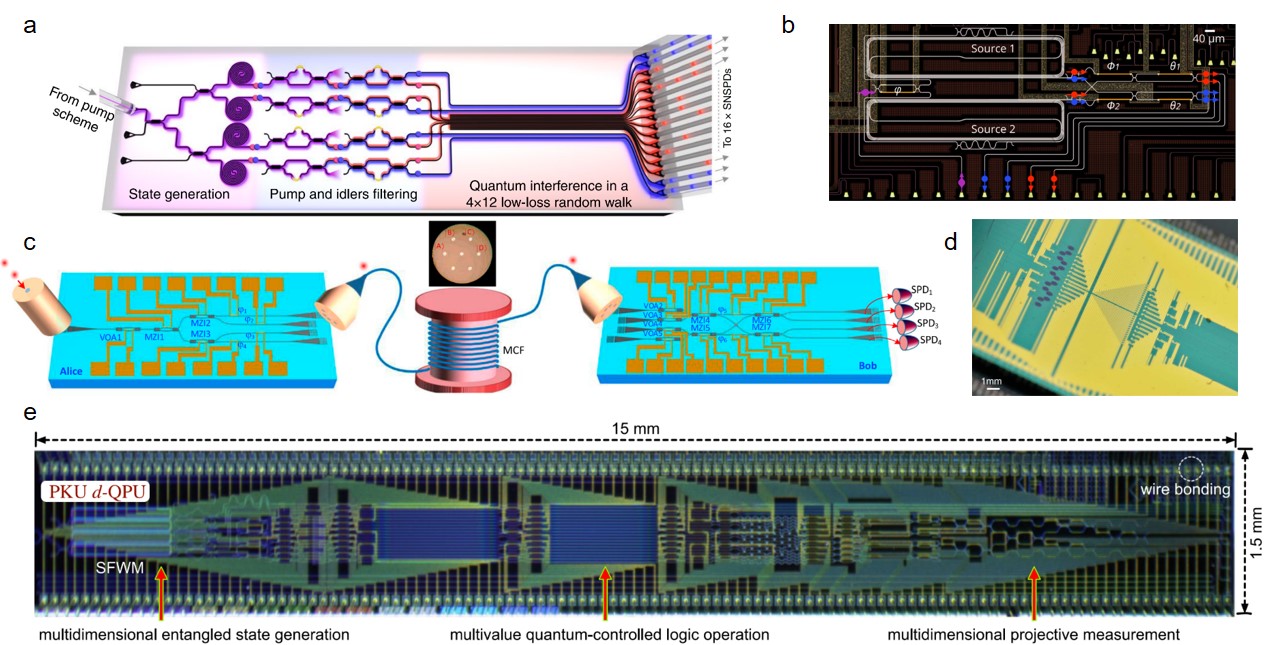}
\caption {\textbf{Multiphoton and high dimensional applications with silicon photonic devices.} a. Silicon photonic chip for the generation and sampling of quantum states. Adapted from Ref. \cite{Paesani2019}. b. Coherent pumping of two sources and processing of the emitted photons. Adapted from Ref. \cite{Paesani2020}. c. Chip-to-chip high-dimensional quantum key distribution based on multicore fiber. Adapted from Ref. \cite{Ding2017}. d. Silicon device for multidimensional quantum entanglement. Adapted from Ref. \cite{Wang2018}. e. Programmable qudit-based quantum processor. Adapted from Ref. \cite{Chi2022}. } 
\label{fig7}
\end{figure*}

Assuming that a system contains $n$ photons with $m$ dimension, the system capacity is $m^n$. Therefore, increasing the number of photons will improve the system capacity exponentially, and it is vitally important. Although multiple solid-state quantum emitters can be integrated on the same chip and multiple determinate single photons can be generated, controlling emitters to ensure indistinguishability between photons remains a major difficulty. Through multiplexing techniques, single emitter can also generate multiple identical photons and complex quantum states \cite{Istrati2020,Wangw2017}, however, there are still difficulties in on-chip integration of multiplexing components. Alternative approach, SFWM can generate multiple photons directly on the single chip and the relevant work have made great progress in recent years.

With strong third-order nonlinear response, silicon waveguides and micro-ring resonators have been used to generate entangled photon pairs with different degrees of freedom \cite{Feng2019,Takesue2007,Takesue2008,Matsuda2012,Silverstone2015,Li2017} and to demonstrate various optical quantum applications \cite{Wangwang2017,Paesani2017,Qiang2018,Qiang2021}. Furthermore, multiple photon pairs were multiplexed to generate quantum states consisted with more photons. In 2011, ref. \cite{Harada2011} has performed quantum interference experiment with two heralded photons generated in two independent silicon waveguides. A visibility of 73\% has been observed. Later, in 2018, this multi-photon interference process has been demonstrated on one single chip with heralded photons from two independent micro-ring resonator sources and interference fringe visibility as 72\% was measured \cite{Faruque2018}. Almost at the same time, silicon waveguides have been used to prepare complex four-photon states \cite{Zhang2019} and frequency-degenerate entangled four-photon states \cite{Feng20192}. These works stimulate more multi-photon applications in integrated optical chips, such as preparation of programmable four-photon graph states \cite{Adcock2019}, generation and sampling of quantum states of light (Fig. 7a) \cite{Paesani2019} and observing nonlocal quantum interference \cite{Feng2021}. So far, the number of photons on one single silicon chip has been increased to 8 \cite{Paesani2019} and more photons could be achieved by further reducing losses. In improving the quality of multi-photon interference, different strategies have been demonstrated to improve the spectral purity of photon pairs \cite{Paesani2020,Liu2020,Burridge2020}. In particular, ref. \cite{Paesani2020} has achieved on-chip heralded two-photon quantum interference with a visibility of 96\% (Fig. 7b). These high-quality integrated photonic sources are promising in practical quantum applications.

High-dimensional encoding is another feasible approach to a larger system capacity. Moreover, it shows many unique quantum properties and provides improvements in particular applications such as higher capacity and noise robustness in quantum communications \cite{Hu2018} and higher efficiency and flexibility in quantum computing \cite{Lanyon2009}. On integrated chips, many degrees of freedom can be used for high-dimensional encoding, such as path, transverse-mode, frequency and time-bins. Among them, path encoding is the most common one due to its ease of implementation. With on-chip beam splitter and Mach-Zehnder interferometers, photons can be routed and manipulated in multiple paths. Coupled waveguide array also can be used to construct the desired dynamic evolution Hamiltonian for research including quantum walk, boson sampling and quantum simulation. 

For high-dimensional quantum applications with integrated devices, ref. \cite{Ding2017} has experimentally demonstrated high-dimensional chip-to-chip quantum interconnection for the first time in 2017 (Fig. 7c). High-dimensional quantum information generated in one silicon photonic chip has been distributed to another chip through one multicore fiber. 
Combined with on-chip quantum photonic sources, programmable high-dimensional bipartite entangled systems have been realized \cite{Wang2018,Lu2020}. 
Recently, a programmable qudit-based quantum processor including all key functions for initialisation, manipulation, and measurement of two quantum quart states and multi-value quantum-controlled logic gates has been demonstrated on a silicon chip (Fig. 7e) \cite{Chi2022}.

Despite path, other degrees of freedom also have advantages in increasing the system capacity. For example, integrated photonic sources always have multiple frequencies \cite{Feng2020} and waveguide transverse-mode shows a compact way for parallel encoding \cite{Dai2018}. Furthermore, it is effective to use multiple degrees of freedom of a quantum particle simultaneously \cite{Feng2016}. The development of on-chip conversion devices will greatly increase the application prospects of these degrees of freedom.

\subsection{B. Quantum error correction}

\begin{figure*}[t]
\centering
\includegraphics[width=13cm]{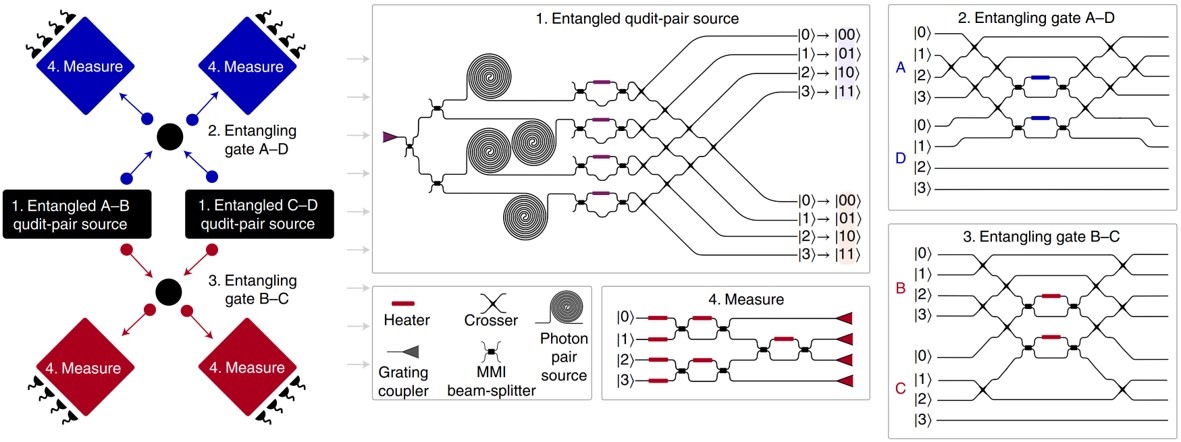}
\caption {\textbf{Quantum error correction with silicon photonic devices.} Error-protected qubits for quantum computation. Adapted from Ref. \cite{Vigliar2021}.} 
\label{fig8}
\end{figure*}

Despite many natural advantages, quantum information processor faces one basic difficulty, that is, the errors. When the product of logical operations and error rates of qubits reaches a certain level, the logical operations will loss their reliability. Meanwhile, due to quantum no-clone theorem, qubits cannot be copied, and eliminate errors by repetition as in classical processors is impossible. Therefore, effective quantum error correction is a necessary step to realize one realistic quantum processor. In photonic quantum technologies, 
photon loss, quality of quantum states, and imperfect logical operations are all sources of errors. Various schemes were developed to improving fault-tolerant capability with photons, such as making efforts for decreasing fabrication imperfections to obtain high-performance optical components. In ref. \cite{Wilkes2016}, 60 dB extinction ratio MZI can be achieved with improved design, which equivalently can implement single-qubit quantum gate with 99.9999\% fidelity. 

Another famous method is the generation of large-scale cluster states for measurement-based quantum computation (MBQC) \cite{Yao2012,Larsen2019,Asavanant2019}, which depends on a sequence of local measurements on a special entangled state, called the cluster or graph state. MBQC is equivalent to the circuit-based quantum computation model, and since it is relatively easy to prepare this kind of entangled states with nonlinear photonics, this model is highly valued in photonic quantum information applications. Besides being used in quantum computation, cluster state also shows important applications in other fields such as quantum error correction, multi-partite quantum communication and quantum metrology, as well as in the study of fundamental problems such as non-locality and decoherence \cite{Hein2006,Shettell2020}. 
So far, with silicon photonic circuits, all types of four-photon graph state have been programmably generated \cite{Adcock2019}, and 
a range of quantum information processing tasks with and without error-correction encodings have been implemented (Fig. 8) \cite{Vigliar2021}. The success rate was increased from 62.5\% to 95.8\% when running a phase-estimation algorithm by using the error-correction program. 
Similar to MBQC, fusion-based quantum computation was proposed and developed in recent years by PsiQuantum \cite{Bartolucci2021,Bombin2021}. With small-scale entangled photons as resource states and fusion-measurement, large-scale universal fault-tolerant quantum computing can be realized with on-chip components.     

Other novel error-correction methods includes entanglement purification and topologically electromagnetic modes. Entanglement purification is a way to extract a subset of states of high entanglement and high purity from a large set of less entangled states, which can significantly increase the quality of logic operations between different qubits and relax the requirement for high-accuracy logic operations \cite{Pan2003}. The topologically electromagnetic modes are much less affected by nanophotonic fabrication-induced disorder and can be used to avoid errors. Up to now, topological quantum light sources \cite{Mittal2018,Wang2019,Dai2022} and quantum interference process \cite{Chenchen2021} on silicon photonic chips have been demonstrated.

\subsection{C. Quantum key distribution}

\begin{figure*}[t]
\centering
\includegraphics[width=13cm]{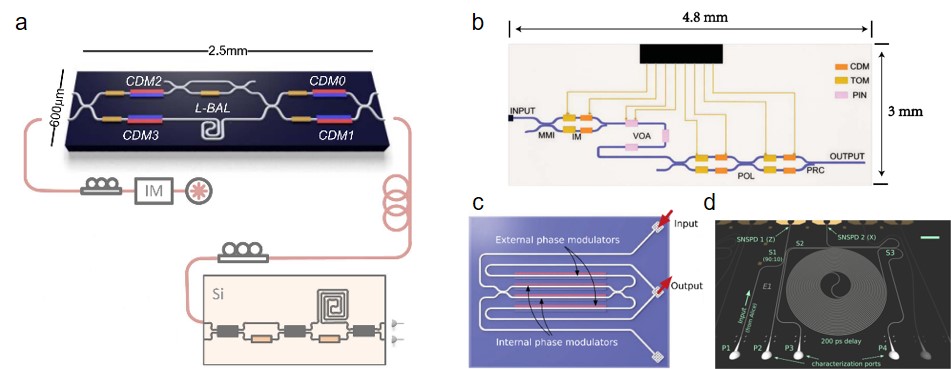}
\caption {\textbf{Quantum key distribution (QKD) with silicon photonic devices.} a. Integrated devices for time-bin encoded BB84. Adapted from Ref. \cite{Sibson2017}.  b. Integrated devices for high-speed measurement-device-independent QKD. Adapted from Ref. \cite{Wei2020}. c. Silicon photonics encoder with high-speed electro-optic phase modulators. Adapted from Ref. \cite{Bunandar2017}. d. Detector-integrated on-chip QKD receiver. Adapted from Ref. \cite{Beutel2021}.} 
\label{fig9}
\end{figure*}

Governed by the laws of quantum mechanics, quantum key distribution (QKD) aims to share information with absolute security between the transmitters and receivers. After decades of development, QKD is becoming the building block of quantum network, and many quantum encoding protocols have been developed, such as BB84, two-state and Einstein-Podolsky-Rosen protocol \cite{Gisin2002}. As the common choice of information carrier, photons are distributed over 803-km fiber \cite{Wangw2022} and from the satellite to the ground over a distance of up to 1,200 kilometres \cite{Liao2017}. For convenience and practicality purposes, transmitters and receivers have been integrated on the photonic chip to achieve dense integration, high stability and scalability \cite{Wang2021}, and are connected by single-mode or multicore fibers \cite{Ding2017,Sibson2017,Bacco2017}. 

Through combining slow thermo-optic DC biases and fast (10 GHz bandwidth) carrier-depletion modulation, high-speed low-error QKD modulation has been achieved with silicon photonic devices \cite{Sibson2017}. Besides, QKD systems with integrated silicon photonics have been demonstrated in an intercity metropolitan test with a 43-km fiber \cite{Bunandar2017} and even full daylight \cite{Avesani2019}. The QKD system is becoming more compact and different degrees of freedom encoding have been implemented such as path \cite{Ding2017}, polarization \cite{Bunandar2017} and time-bins \cite{Beutel2021,Geng2019}. Based on hybrid techniques, photon lasers \cite{Agnesi2019} and photon detection processes \cite{Rafaelli2018,Beutel2021,Zheng2021} have been integrated directly on the silicon chip. The high-dimensional QKD with quantum states generated in the silicon photonic circuits enables surpassing the information efficiency limit of traditional quantum key distribution protocols \cite{Ding2017}. With these technological advances, different key distribution protocols have been achieved with integrated silicon photonic devices such as BB84 \cite{Sibson2017}, continuous-variable \cite{Zhangzhang2019} and measurement-device-independent \cite{Wei2020}, as shown in Fig. 9.

\subsection{D. Quantum state teleportation}

\begin{figure*}[t]
\centering
\includegraphics[width=13cm]{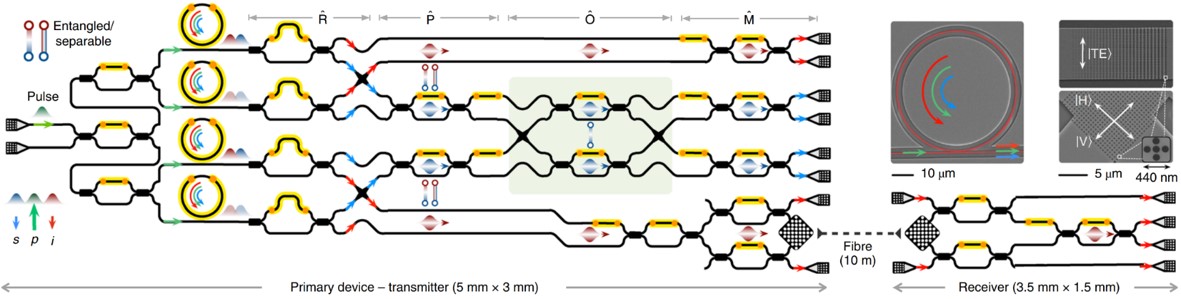}
\caption {\textbf{Quantum state teleportation with silicon photonic devices.} Chip-to-chip quantum teleportation. Adapted from Ref. \cite{Llewellyn2020}.} 
\label{fig10}
\end{figure*}

Quantum state teleportation aims to transfer a particle's quantum state to another particle rather than itself. As a new way of communication, this technique transfers quantum information carried in the quantum state with the help of quantum entanglement, and forms the basis of scalable quantum network and distributed quantum computing. It involves three particles (A, B, C), and we assume particle B and C are Bell entangled. The quantum state of particle A can be transferred to particle B through Bell state measurement between A and C. With quantum state teleportation, linear optical quantum information processor can be efficiently scaled by using a large number of cascaded gates \cite{Kok2007}. 

The integrated quantum teleportation process was first implemented with silica slab waveguides \cite{Metcalf2014}, 
which was fabricated by the direct UV-writing technique. The silica waveguides have large cross-section (typically 4.5$\upmu$m$\times$4.5$\upmu$m), and low chip-fiber coupling loss and transmission loss enable us to achieve single photons generated in free-space feeding into the chip directly through optical fibers while maintaining high brightness. In ref. \cite{Metcalf2014}, four photons which were generated in free-space nonlinear crystals were used. Three photons were input into the chip, and the left one was input into the detector as heralding. On-chip interferometers were used to realize the functions of entanglement generation and Bell state measurement. Based on state measurement results, classical communications and corresponding quantum state operations, quantum state of one photon would be transferred to another photon. 

Of course, quantum teleportation process can be achieved with silicon waveguides. In particular, strong SFWM in silicon waveguides can be used directly to manipulate multi-photon entangled quantum states. In ref. \cite{Llewellyn2020}, quantum teleportation between two silicon chips have been demonstrated, as shown in Fig. 10. Four microresonators were used to generate high-quality entangled quantum states directly on the silicon chip. Before the validation of teleportation process, photons that obtained the quantum states were transmitted to another chip with quantum photonic interconnect by path-polarization interconversion \cite{Wang2016}, where quantum states were reconstructed via state tomography measurements. This achievement 
lays the foundation for large-scale integrated photon quantum technology in communication and computing.

\section*{IV. Challenges and outlook}

Despite progress mentioned above, further improvements are needed in some areas for scalable quantum information applications. Below, we outline some challenges in silicon photonic quantum technologies.

\textbf{Low-Loss Components:} Loss is one huge challenge for optical quantum integrated systems, including loss of passive structures, delay lines, switches and chip interconnects. Because of the small effective mode area, loss in silicon is more prominent than in other materials, such as silica. Although the state-of-the-art low transmission loss of 0.08 dB/m has been demonstrated based on sidewall smoothing
techniques \cite{Lee2012}, it is difficult to apply to waveguide structures with common strip waveguides. Multimode waveguide technology and mixing with silica or silicon nitride could enable the next generation of ultra low-loss components.    

\textbf{Photon Generation:} For parametric photonic sources, the silicon waveguide has strong two-photon absorption at 1550 nm, and it is difficult to enhance the brightness of the photonic sources just through increasing the pump power. Besides, the properties of the photonic sources are relatively poor, especially for multi-photon interference. Photonic source capable of large-scale expansion with high brightness is still on the way. Extending photons to the far infrared band, where silicon has low two-photon absorption, is one potential method. Another possible solution is to develop hybrid integrated silicon chips with other materials with better nonlinear properties, such as silicon nitride, in which preparation of tens of photons has been demonstrated \cite{Arrazola2021}, and lithium niobate, which has strong second-order nonlinear response. The stochastic character of the photon sources needs to be solved, and multiplexed photon sources, long-time delay and high-speed modulation should be integrated totally on the same chip. For deterministic single-photon sources, modulation to increase the indistinguishability between different sources should be developed.    

\textbf{Deterministic Quantum Operation:} The negligible photon-photon interaction limits the applicability of many photon-based schemes. To overcome this limitation, multiplexed photon sources and feedforward capability based on heralded detection of auxiliary photons should be integrated totally on the same chip. Strong nonlinear media such as atoms need to be introduced into the chip to enhance the photon-photon interaction to build deterministic multi-photon gates.

\textbf{Frequency Conversion:} Efficient frequency conversion will link various quantum systems to build up quantum networks, such as photonic conversion between microwave and telecom C-band. Efficient conversion at the single-photon level will allow us to take advantage of different systems, even if they are far away. 

We have witnessed tremendous advances in silicon photonic devices for quantum information processing, especially in recent years. Photon source, quantum state manipulation and detection can be integrated on a chip and hopefully on the same chip, and integrated programmable multi-photon and high-dimensional quantum information processors have been demonstrated. With the further upgrading of fabrication technology, silicon photonics will have a greater prospect in quantum information processing. Of course, silicon photonic devices still face many defects of the material itself, and the future quantum information processors are most likely to be hybrid with various materials elevated to the extreme. In any case, we believe that silicon photonics will play an important role.

\textbf{Funding.} This work was supported by the National Natural Science Foundation of China (NSFC) (Nos. 62061160487, 62005239, 12004373, 61974168, 62075243), the Innovation Program for Quantum Science and Technology (No. 2021ZD0303200), the Natural Science Foundation of Zhejiang Province (LQ21F050006), the National Key Research and Development Program (2017YFA0305200), the Key Research and Development Program of Guangdong Province of China (2018B030329001 and 2018B030325001), the Postdoctoral Science Foundation of China (No. 2021T140647) and the Fundamental Research Funds for the Central Universities. This work was partially carried out at the USTC Centre for Micro and Nanoscale Research and Fabrication.

\textbf{Disclosures.} The authors declare no conflicts of interest.





\end{document}